\documentstyle[preprint,aps,floats]{revtex}

\begin{document}

\title{Non-Transitive Quantum Games}

\author{Michael L. Stohler$^{1,2}$ and Ephraim Fischbach$^{2}$}

\address{$^{1}$Physics Department, Wabash College, Crawfordsville, IN
47933-0352}
\address{$^{2}$Physics Department, Purdue University, West Lafayette, IN
47907-1396}

\date{\today}

\maketitle

\begin{abstract}
Non-transitivity can arise in games with three or more strategies $A,B,C$, when $A$ beats $B$, $B$ beats $C$, and $C$ beats $A$,
($A>B>C>A$).  An example is the children's game \textquotedblleft rock, scissors, paper" ($R,S,P$) where $R>S>P>R$.  We discuss the
conditions under which quantum versions of
$R,S,P$ retain the non-transitive characteristics of the corresponding classical game.  Some physical implications of
non-transitivity in quantum game theory are also considered.
\end{abstract}

%\begin{multicols}{2}

\pagebreak

Recent years have witnessed a rapid growth of interest in quantum game theory, motivated in part by potential applications to quantum computing. 
Quantum games are generally derived from the corresponding classical games by introducing some inherently quantum
mechanical feature (such as superposition of states\cite{meyer} or entanglement\cite{mw}), which can be incorporated in more than one way.  In a 2-player
game, where each player has two pure strategies available, a widely discussed scheme for entanglement is that due to Eisert, Wilkins, and Lewenstein (EWL)
\cite{ewl}.  The object of the present paper is to discuss 2-player games with three pure strategies ($A,B,C$) using the EWL formalism for entanglement. 
The novel feature of such games is that they allow for the possibility of non-transitive strategies ($A>B>C>A$).  It is well known that non-transitivity
arises in classical games
\cite{game} and in real-world applications, and can lead to surprising -and seemingly paradoxical - outcomes.  This naturally
leads to the question of whether similar non-transitive effects arise in quantum versions of classical games and, if so, whether similarly unexpected
effects can be present at the quantum level.

A simple 2-player, 3-strategy game is the children's choosing game \textquotedblleft rock ($R$), scissors ($S$), paper ($P$)" in which rock beats scissors,
scissors beats paper, and paper beats rock ($R>S>P>R$).  A payoff matrix for the 3 strategies of this zero-sum game is shown in Table I.  (Quantum games
with $3 \times 3$ payoff matrices and larger have have been discussed by Wang, et al. \cite{wang}.)  We can formulate a quantum analog of this game by
using the EWL entanglement formalism.  In the quantum version of this game the strategies
$R$,
$S$, and $P$ are represented by three matrices which act on qubits from which payoffs are determined.  The quantum game can be played as follows:  In the
absence of entanglement each player (Alice and Bob) is given a separate copy of the same qubit which is in one of three orthogonal states denoted by
$|100\rangle$,
$|010\rangle$, or $|001\rangle$. Without loss of generality we assume that the same qubit is used for the initial state in every game.  This qubit can
then be manipulated by any of three matrices denoted by $R$, $S$, and $P$, each of which rotates the initial state into one of the three orthogonal
states.  Once both players have implemented their strategies ($R$,
$S$, or $P$), the final qubits for both players are compared using the payoff matrix in Table I.  An examination of this Table shows that
$R>S>P>R$.  In the EWL formalism the qubits are entangled by an operator before the players act on them with $R$, $S$, or $P$, and then disentangled
afterwards.  Since the EWL entanglement operator commutes with any direct-product combination of $R$, $S$, and $P$, the net effect of
introducing entanglement in this manner is to produce a quantum game whose outcomes are identical to those of the corresponding
classical game.

If a game is repeated many times, a competitor may elect to play any of $R,S,$ or $P$ in each game, and in such a circumstance quantum games can be
formulated in which payoff functions unique to quantum mechanics may result.  If the probabilities for the actions
$R$,
$S$, or
$P$ are
$p_R$, $p_S$, and $p_P$ respectively, a player is said to be adopting a {\it mixed strategy}.  Two ways of incorporating mixed strategies in quantum games
can be considered, and these lead in general to different outcomes.  In one approach, $p_R$, $p_S$, and $p_P$ are simply the classical probabilities of
using each of the quantum operators
$R$,
$S$, or
$P$.  By contrast, the second approach combines the operators $R$, $S$, or $P$ and the respective probability amplitudes into a single matrix.  The difference in
the two constructions can be illustrated in the $2 \times 2$ case (Alice's $2$ choices $\times$ Bob's $2$ choices): In the former approach each player would have
only two options ($N=no-flip$ and
$F=flip$) for each game,

\begin{equation}
N=\left(
\begin{array}{cc}
1 & 0 \\
0 & 1  
\end{array}
\right),\;\;\;
F=\left(
\begin{array}{cc}
0 & 1 \\
1 & 0  
\end{array}
\right),
\label{j}
\end{equation}
which act on the initial game state, either $(10)$ or $(01)$.  Assigning a classical probability $p$ to using $N$ and $(1-p)$ to using $F$ would then
define a mixed strategy.  It should be noted that other choices for $N$ and $F$ are interesting to study\cite{ew}, but we shall restrict our discussion to
those given in Eq. (\ref{j}).  This entanglement technique requires that a player make a choice in each game.  A maximized entanglement operation, $J$,
can be introduced into such a game that commutes with any direct-product combination of
$N$ and
$F$ but does not commute with a general matrix,

\begin{equation}
J=\frac{1}{\sqrt{2}}(N \otimes N)+\frac{i}{\sqrt{2}} (F \otimes F),
\label{k}
\end{equation}
where $N \otimes N=N(Alice) \otimes N(Bob)$, etc.. Payoff functions for both players unique to quantum mechanics are possible if either player cheats by
using neither
$N$ nor
$F$.
 
In the second approach the actions available to each player, along with the probability {\it amplitudes} for
selecting them, are built into the single unitary matrix
$U$, 

\begin{equation}
U=\left(
\begin{array}{cc}
\sqrt{p} & -\sqrt{1-p} \\
\sqrt{1-p} & \sqrt{p}  
\end{array}
\right)=\left(
\begin{array}{cc}
\cos{\theta} & -\sin{\theta} \\
\sin{\theta} & \cos{\theta}  
\end{array}
\right),
\label{l}
\end{equation}
Once $p=p(\theta)$ has been chosen, a game can be played many times without the player
making any future decisions. An entanglement matrix for such a game was provided by EWL,

\begin{equation}
J=\frac{1}{\sqrt{2}}(N \otimes N)+\frac{i}{\sqrt{2}} (F' \otimes F'),
\label{m}
\end{equation}
where

\begin{equation}
F'=\left(
\begin{array}{cc}
0 & -1 \\
1 & 0  
\end{array}
\right).
\end{equation}

Before turning to the 3-strategy case we note that although some properties of quantum games are identical to those of classical games\cite{enk}, the EWL
entanglement technique can lead to payoff functions that cannot be reproduced in a classical game.  Consider a situation where each player introduces an
additional phase,
$\phi$, to his/her play:
\begin{equation}
U(\theta, \phi)=\left(
\begin{array}{cc}
e^{i \phi} \cos{\theta} & -\sin{\theta} \\
\sin{\theta} & e^{-i \phi} \cos{\theta}  
\end{array}
\right)
\label{n}
\end{equation}The final expected payoff function, $\bar{\$}_a$, for Alice is

\begin{eqnarray}
\bar{\$}_a &=&a_{11}[c_a^2 c_b^2 \cos^2{(\phi_a+\phi_b)}]\\
&&+a_{10}(c_a s_b \cos{\phi_a}-s_a c_b \sin{\phi_b})^2  \nonumber \\
&&+\mbox{} a_{01}(s_a c_b \cos{\phi_b}-c_a s_b \sin{\phi_a})^2 \nonumber \\
&&+a_{00}[s_a s_b + c_a c_b \sin{(\phi_a+\phi_b)}]^2,  \nonumber 
\end{eqnarray}
where $\phi_a(\phi_b)$ is the additional phase for Alice(Bob), $c_a=\cos{\theta_a}$, $s_a=\sin{\theta_a}$ and the $'a'$s are Alice's payoff
coefficients.  For non-zero but fixed values of
$\phi_a$ and
$\phi_b$ Alice's payoff is non-linear in
$p_a=c_a^2$,  whereas a classical $2 \times 2$ game produces only payoff functions that are linear in $p_a$.

The $2 \times 2$ EWL entanglement formalism can be easily shown to work equally well with either formulation of a mixed strategy.  However, in the
$3
\times 3$ case (2 players with 3 choices each), these approaches require different entanglement matrices as we now discuss.  We begin by exhibiting the $3
\times 3$ analog of the operators $N$ and $F$ in Eq. (\ref{j}) which we take to be:

\begin{equation}
U_1=\left(
\begin{array}{ccc}
1 & 0 & 0  \\
0 & 1 & 0  \\
0 & 0 & 1 
\end{array}
\right),\;\;\;
U_2=\left(
\begin{array}{ccc}
0 & 0 & 1 \\
1 & 0 & 0 \\ 
0 & 1 & 0 
\end{array}
\right),\;\;\;
U_3=\left(
\begin{array}{ccc}
0 & 1 & 0 \\
0 & 0 & 1 \\ 
1 & 0 & 0 
\end{array}
\right),
\end{equation} where $[U_2,U_3]=0$. The players use these matrices in different spaces to act on initial wave functions such as $|11
\rangle=(100)_a\otimes(100)_b$.  The analog of the first
approach for constructing a mixed strategy is for each player to assign classical probabilities $p_i$ to each of the
$U_i$ where
$
\sum p_i=1$. (Other choices for these matrices lead to identical results.)  The commutativity of $U_1$, $U_2$ and $U_3$ can be exploited to create a
mixing matrix similar to that of EWL\cite{ewl}.

In direct analogy to Eq. (\ref{k}), a choice for the unitary entanglement matrix, $J$, that commutes with any
direct product combination of these three matrices is:

\begin{eqnarray}
J&=&\frac{-1}{3} U_1 \otimes U_1+\frac{2}{3} U_2 \otimes U_2+ \frac{2}{3} U_3 \otimes U_3   
\label{p}
\\
&=&\left(
\begin{array}{ccccccccc}
\frac{-1}{3} & 0 & 0 & 0 & \frac{2}{3} & 0 & 0 & 0 & \frac{2}{3}  \\
0 & \frac{-1}{3} & 0 & 0 & 0 & \frac{2}{3} & \frac{2}{3} & 0 & 0  \\
0 & 0 & \frac{-1}{3} & \frac{2}{3} & 0 & 0 & 0 & \frac{2}{3} & 0  \\
0 & 0 & \frac{2}{3} & \frac{-1}{3} & 0 & 0 & 0 & \frac{2}{3} & 0  \\
\frac{2}{3} & 0 & 0 & 0 & \frac{-1}{3} & 0 & 0 & 0 & \frac{2}{3}  \\
0 & \frac{2}{3} & 0 & 0 & 0 & \frac{-1}{3} & \frac{2}{3} & 0 & 0  \\
0 & \frac{2}{3} & 0 & 0 & 0 & \frac{2}{3} & \frac{-1}{3} & 0 & 0  \\
0 & 0 & \frac{2}{3} & \frac{2}{3} & 0 & 0 & 0 & \frac{-1}{3} & 0  \\
\frac{2}{3} & 0 & 0 & 0 & \frac{2}{3} & 0 & 0 & 0 & \frac{-1}{3}  \\
\end{array}
\right),  \nonumber 
\end{eqnarray}
where $U_1 \otimes U_1 \equiv U_1(Alice) \otimes U_1(Bob)$, etc.. The entanglement matrix in Eq. (\ref{p}) is designed to be used in a game where players
randomly select an action for each game based on classical probabilities associated with a defined mixed strategy.  This matrix entangles the qubits while
still obeying the appropriate commutation relations,

\begin{equation}
0=\left[ J, U_i \otimes U_j \right] \;\;\;\;\;\;\;\;\;\;\;\forall\;\;i,j.
\end{equation}Unlike the $2 \times 2$ case, an initial eigenstate does not become fully entangled by this matrix.  The initial state
$|11\rangle$, for example, is transformed into

\begin{equation}
J|11\rangle=-\frac{1}{3}|11\rangle+\frac{2}{3}|00\rangle+\frac{2}{3}|-1-1\rangle.
\end{equation}
If each amplitude is denoted as $c_i$, then the degree of entanglement (as measured by von Neumann entropy\cite{entropy}), $E$, is:
\begin{equation}
E=- \sum_{i} |c_i|^2\log_3{|c_i|^2}
=0.88
\end{equation}
where log base 3 is used to ensure that maximum entanglement corresponds to $E=1$.

A generic entanglement matrix can be constructed for an $N \times N$ game.  Let $U_1$,$U_2$,...$U_{N}$ ($U_iU_j=U_{i+j}$, $U_i^\dag=U_{-i}$ and
$U_{N+1}=U_1$) be commuting and orthogonal matrices, each capable of directly transforming an initial state into an eigenstate of some specified
Hamiltonian.  The elements of each matrix are taken to be either $0$ or $1$.  If $J$ is assumed to have the form

\begin{equation}
J=\sum_{i=1}^{N} \alpha_{i}U_i \otimes U_i,
\label{zero}
\end{equation}
then

\begin{eqnarray}
JJ^{\dag}&=&\sum_{i=1}^{N} \alpha_{i} U_i \otimes U_i \sum_{j=1}^{N} \alpha_{j}^*U_j^\dag \otimes U_j^\dag,
\\
&=&\openone \otimes \openone \sum_{k=1}^{N} |\alpha_{k}|^2+\mathop{\sum_{i=1}^{N}}_{i \neq j} \sum_{j=1}^{N} \alpha_i \alpha_j^* U_iU_j^\dag \otimes
U_iU_j^\dag.
\end{eqnarray}
Unitarity requires that (1) 

\begin{equation}
1=\sum_{k=1}^{N} |\alpha_{k}|^2,
\label{one}
\end{equation}
and (2) that

\begin{equation}
0=\sum_{i=1}^{N} \sum_{j=0}^{N-1} \alpha_i \alpha_j^* U_iU_j^\dag \otimes
U_iU_j^\dag.  
\end{equation}
If $s \equiv i-j$, then the second unitarity condition becomes

\begin{eqnarray}
0&=&\sum_{j=1}^{N} \alpha_{s+j} \alpha_j^* U_s \otimes U_s \;\;\;\;\;\;\;\;\;\;\; \forall \; s[1,N-1],
\\
&=&U_s \otimes U_s\sum_{j=1}^{N} \alpha_{s+j} \alpha_j^*  \;\;\;\;\;\;\;\;\;\;\; \forall \; s[1,N-1],
\end{eqnarray}
or

\begin{equation}
0=\sum_{j=1}^{N} \alpha_{s+j} \alpha_j^*  \;\;\;\;\;\;\;\;\;\;\; \forall \; s[1,N-1].
\label{two}
\end{equation}
For the $N=3$ case, $\alpha_1=-1/3$, $\alpha_2=\alpha_3=2/3$ is one solution to the the unitarity requirements.  For the general $N \times N$ case, the
combination of Eqs. (\ref{one}) and (\ref{two}) thus leads to an expression for the $\alpha_i$ in Eq. (\ref{zero}), and hence to an appropriate
expression for $J$.  

 The $3 \times 3$ game that we consider is \textquotedblleft rock, scissors, paper" (R,S,P) shown in Table I which has the
interesting non-transitive property ($R>S>P>R$).  As shown in the Table, a payoff of
$+1$ has been assigned to winning, $-1$ to losing, and $0$ for both in case of a tie. It is well known that in a $2 \times 2$
game such as the \textquotedblleft prisoner's dilemma", a player using a quantum strategy can improve his/her expected payoff provided that
his/her opponent continues to use a classical strategy.  The question then arises whether this can also happen in a $3 \times
3$ game which is non-transitive at the classical level. To address this question we note that the generic matrix, $\hat{J}$, 

\begin{eqnarray}
\hat{J}&=&\alpha_1 U_1 \otimes U_1+\alpha_2 U_2 \otimes U_2+ \alpha_3 U_3 \otimes U_3  
\label{q}
\\
&=&\left(
\begin{array}{ccccccccc}
a & 0 & 0 & 0 & c & 0 & 0 & 0 & b  \\
0 & a & 0 & 0 & 0 & c & b & 0 & 0  \\
0 & 0 & a & c & 0 & 0 & 0 & b & 0  \\
0 & 0 & b & a & 0 & 0 & 0 & c & 0  \\
b & 0 & 0 & 0 & a & 0 & 0 & 0 & c  \\
0 & b & 0 & 0 & 0 & a & c & 0 & 0  \\
0 & c & 0 & 0 & 0 & b & a & 0 & 0  \\
0 & 0 & c & b & 0 & 0 & 0 & a & 0  \\
c & 0 & 0 & 0 & b & 0 & 0 & 0 & a  \\
\end{array}
\right)  \nonumber 
\end{eqnarray}
has no effect on the payoff
matrix because it transforms eigenstates to others with identical payoffs.  It follows from this discussion that the matrix $J$ in Eq. (\ref{p}), which
satisfies all the criteria for a mixing matrix in $3 \times 3$ games, produces outcomes for the quantum version of the non-transitive game $R,S,P$ which
are identical to those for the classical game.  Since this conclusion also holds for the generic matrix $\hat{J}$ in Eq. (\ref{q}) (assuming unitarity),
we can conclude that the quantum versions of $R,S,P$ constructed from $J$ or
$\hat{J}$ will retain the non-transitive behavior present in the classical game.  We emphasize, however, that the application of $J$ or $\hat{J}$ to other
$3 \times 3$ games can be expected to produce quantum games whose outcomes are generally different from those of the corresponding classical game.

Since the survival of non-transitivity in the quantum domain is a fundamental feature in $3 \times 3$ games, we present an alternative proof of this
result which clarifies some of the underlying assumptions.  We assume that Alice uses a strategy (which is separable into a product of
'classical' and 'quantum' operators), $Q'=U_iQ$, for every game and that Bob uses some 'classical' strategy $U_j$.  In this case, the final game state is

\begin{equation}
| \psi_f \rangle = J_b^{\dag} [U_iQ \otimes U_j] J_b |11 \rangle.
\end{equation}
 If $| \psi_{jq} \rangle \equiv J_b^{\dag} [Q \otimes \openone] J_b |11 \rangle $ then the final state becomes

\begin{equation}
| \psi_f \rangle = U_i \otimes U_j | \psi_{jq} \rangle.
\end{equation}
Non-transitivity survives because Bob is still able to win regardless of what Alice does: he can turn any advantage that Alice creates with $Q$
into an advantage to himself simply by choosing another $U_j$.  It is important to note that, unlike the \textquotedblleft prisoner's dilemma", the use of
quantum strategies by both players cannot be mutually beneficial to both players because the game is zero sum.  

In the preceding discussion we considered non-transitive $3 \times 3$ games in a framework analogous to that following Eq. (\ref{j}) in the $2 \times 2$
case.  We turn next to a quantum formulation of $3 \times 3$ games analogous to that following from Eq. (\ref{l}).  It can be shown that the preceding
proof of non-transitivity survives for this formulation as well.  The matrix that can be used to transform an initial
eigenstate into any other eigenstate is generated by the following matrices, which are Hermitian and unitary,

\begin{equation}
H_1=\left(
\begin{array}{ccc}
1 & 0 & 0 \\
0 & 0 & 1 \\ 
0 & 1 & 0 
\end{array}
\right), \;\;\;
H_2=\left(
\begin{array}{ccc}
0 & 1 & 0 \\
1 & 0 & 0 \\ 
0 & 0 & 1 
\end{array}
\right).
\end{equation}

Each player can use the matrix $U(x,y)$,
\begin{eqnarray}
U(x,y)&=&e^{ixH_1} e^{iyH_2}
\\
&=&\left(
\begin{array}{ccc}
e^{ix} \cos{y} & ie^{ix} \sin{y} & 0 \\
i \sin{y} \cos{x} & \cos{x} \cos{y} & i e^{iy} \sin{x} \\ 
-\sin{y} \sin{x} & i \sin{x} \cos{y} & e^{iy} \cos{x} 
\end{array}
\right)
\end{eqnarray}to transform an initial state, $|1 \rangle=(100)$, into any other state to simulate a pure strategy.  For this choice of $U(x,y)$ the
additional phases that have been introduced have no net effect on probabilities or payoff functions.  The choices of
$x$ and
$y$ define a player's mixed strategy in a $3
\times 3$ game, just as the choice of $\theta$  [Eq. (\ref{l})] does in the $2 \times 2$ game.  $U(x,y)$ satisfies the commutation relation
$0=[U(x,y),F'']$, where

\begin{equation}
F''=\frac{1}{3}\left(
\begin{array}{ccc}
-1 & 2 & 2 \\
2 & -1 & 2 \\ 
2 & 2 & -1 
\end{array}
\right).
\end{equation}Because $F''$ and $U(x,y)$ commute, so do $F'' \otimes F''$ and $U(x_a,y_a) \otimes U(x_b,y_b)$, where $x_a$ and $y_a$ ($x_b$ and $y_b$)
define Alice's (Bob's) mixed strategy.  This suggests the introduction of the following (maximally entangled) mixing matrix,

\begin{equation}
\bar{J}=e^{-i \pi/4 F'' \otimes F''}=\frac{1}{\sqrt{2}}[ \openone \otimes \openone - i F'' \otimes F'']
\label{jbar}
\end{equation} which commutes with any $U(x_a,y_a) \otimes U(x_b,y_b)$.  The matrix $\bar{J}$ in Eq. (\ref{jbar}) is the exact $3 \times 3$ analog of the
entanglement matrix used by EWL in their discussion of the
\textquotedblleft prisoner's dilemma"\cite{ewl}.  There is considerable freedom on the part of each player in how he/she chooses to deviate from $U(x,y)$,
since a variable equivalent to $\phi$ can be introduced into a player's transformation matrix in numerous ways.

In summary, we have developed a formalism for dealing with entanglement in $3 \times 3$ quantum games, when each player adopts a mixed strategy.  Although
this formalism can be applied to any
$N
\times N$ game, our focus in this paper has been on the simplest non-transitive
$3
\times 3$ game.  As noted above, classical non-transitive games are of great interest, since they can lead to seemingly paradoxical results in real-world
examples\cite{game}.  One of the central results of the present paper is that non-transitivity survives in the quantum versions of the corresponding
games.  This naturally raises the question of whether similar apparent paradoxes can arise in physically realizable quantum systems.  Although this
question cannot be answered definitively at the present time, physical systems exist which exhibit the non-transitive features of both the
\textquotedblleft voter's paradox" and the \textquotedblleft Penney paradox"\cite{game}.  As we discuss elsewhere\cite{us}, tables of Clebsch-Gordan
coefficients have properties similar to those of the magic square, whose non-transitive properties underlie the \textquotedblleft voter's paradox".  It is
thus possible that interesting non-transitive effects may arise in ensembles of particles with non-zero angular momenta whose behavior can be modeled by
the $3 \times 3$ non-transitive games that we have presented here.

%\acknowledgments
The authors wish to thank Dennis Krause for many helpful discussions.  The work of M.L.S was supported in part by the Wabash College Byron K. Trippet
Research Fund and the work of E.F. was supported in part by the U.S. Department of Energy under Contract No. DE-AC 02-76ER01428.

\pagebreak

%\end{multicols}
\pagebreak

\begin{table}
\caption{In the zero-sum game of \textquotedblleft rock, scissors, paper", a player can win regardless of the strategy chosen by an opponent.  The first
number in each entry of the Table is Alice's payoff and the second is Bob's.  Winning strategies are non-transitive in that $R>S>P>R$.}
\label{isotope table}
\begin{tabular}{lcccc}
       &			 			   &               &      Bob       &      					       \\
       &			 		    &						$R$      &				   $S$      &						  $P$						\\ \hline
       &			$R$				&					$(0,0)$		&	   	$(1,-1)$     &				 	$(-1,1)$				       		\\
Alice		&   $S$				&    	$(-1,1)$  &    	$(0,0)$     &						$(1,-1)$			        		\\
							&   $P$				&     $(1,-1)$  &     $(-1,1)$     &						$(0,0)$				\\
\end{tabular}
\end{table}

\end{document}